\documentclass[manuscript]{aastex61}

\usepackage{graphicx}
\usepackage{makecell}

\newcommand\aastex{AAS\TeX}

\shorttitle{\aastex\ Luminous Efficiency -II}
\shortauthors{Subasinghe, D. \& Campbell-Brown, M.}

\begin{document}

\title{Luminous efficiency estimates of meteors -II. Application to Canadian Automated Meteor Observatory meteor events}


\correspondingauthor{Dilini Subasinghe}
\email{dsubasi@uwo.ca}

\author{Dilini Subasinghe}
\affil{Department of Physics \& Astronomy, University of Western Ontario \\
1151 Richmond Street
London, Ontario, Canada N6A 3K7} 

\author{Margaret Campbell-Brown}
\affil{Department of Physics \& Astronomy, University of Western Ontario \\
1151 Richmond Street
London, Ontario, Canada N6A 3K7}
 
\begin{abstract}
Luminous efficiency is a necessary parameter for determining meteoroid mass from optical emission. Despite this importance, it is very poorly known, with previous results varying by up to two orders of magnitude for a given speed. We present the most recent study of luminous efficiency values determined with modern high-resolution instruments, by directly comparing dynamic and photometric meteoroid masses. Fifteen non-fragmenting meteoroids were used, with a further five clearly fragmenting events for comparison. Twelve of the fifteen non-fragmenting meteoroids had luminous efficiencies less than 1\%, while the fragmenting meteoroids had upper limits of a few tens of percent. No clear trend with speed was seen, but there was a weak negative trend of luminous efficiency on meteoroid mass, implying that smaller meteoroids radiate more efficiently.

\end{abstract}

\keywords{meteorites, meteors, meteoroids,  methods: observational}



\section{Introduction}
Meteoroid masses are poorly constrained. Various studies have used different experimental and observational techniques in the past to determine meteoroid masses from the light they emit, but results vary by up to two orders of magnitude for a given meteoroid speed. This can have consequences for meteoroid flux estimates, which can affect satellites and spacecraft in orbit around Earth \citep{Council2011}. Because all small meteoroids ablate completely in the atmosphere, mass estimates need to be made based on observations which are typically less than one second long. There are many unknown parameters which affect calculations of the mass of a meteoroid, such as its shape and density. Assuming the meteoroid is a solid, non-fragmenting object, its mass can be determined with either of a pair of coupled differential equations that describe the mass loss and deceleration of the meteoroid. The luminous intensity equation (shown in Equation \ref{eq:luminous}) allows one to determine the mass of a meteoroid $m$ from the change in its kinetic energy $E_k$, given the speed $v$ and brightness $I$.

\begin{equation}
I =-\tau\frac{{\rm d}E_{k}}{{\rm d}t} = -\tau\left(\frac{v^{2}}{2}\frac{{\rm d}m}{{\rm d}t} + mv\frac{{\rm d}v}{{\rm d}t}\right)
\label{eq:luminous}
\end{equation} 

A value must be chosen for the proportionality constant $\tau$, the luminous efficiency - a measure of how much of the kinetic energy lost by the meteoroid is used for visible light production. The mass given in Equation \ref{eq:luminous} is the \textit{photometric} mass and includes the mass of all fragments if the meteoroid has broken up. There is a large uncertainty associated with the photometric mass due to the uncertainty in the luminous efficiency. \newline

The deceleration of the meteor is described by Equation \ref{eq:drag}, the drag equation, derived from conservation of momentum.
\begin{equation}
\label{eq:drag}
\frac{\mathrm{d}v}{\mathrm{d}t} = -\frac{\Gamma \rho_{atm} v^2 A}{m^\frac{1}{3}{\rho_m^\frac{2}{3}}}
\end{equation}

It provides a second way to determine the meteoroid mass. In this equation, the mass $m$ is the \textit{dynamic} mass and describes the leading (usually largest and brightest) fragment; $\Gamma$ is the drag coefficient and describes the efficiency of momentum transfer between the atmosphere and meteoroid; $\rho_{atm}$ is the atmospheric density; $\rho_{m}$ is the meteoroid density; and $A$ is the shape factor, defined as the cross sectional area\footnote{mistakenly given as the surface area in \cite{Subasinghe2017}} divided by the object volume to the exponent 2/3. Often, the drag coefficient, the meteoroid density, and the shape factor are assumed to be constant.\newline

Because most meteoroids fragment \citep{Subasinghe2016}, it is more practical to determine the meteoroid mass through the luminous intensity equation, assuming a suitable value can be found for the luminous efficiency.  \newline

The uncertainty in the luminous efficiency is large since it may depend on many factors: meteoroid speed and height; meteoroid and atmospheric composition; the spectral response of the detector; and possibly mass \citep{Ceplecha1998}. Whether each factor has an effect, and its effect on the luminous efficiency, is unknown.  \newline

The luminous efficiency of meteors has been investigated with numerous methods in the past. A theoretical approach was first taken by \cite{Opik1933}, but has been disregarded by many researchers due to theoretical considerations (such as not knowing how quantum states are populated) \citep{Thomas1951,Verniani1965}, and a modern theoretical approach was taken by \cite{Jones2001}, using excitation cross-sections to predict the light produced (however they assumed that ionised atoms are not available for excitation). \cite{Verniani1965} determined the luminous efficiency by equating the dynamic and photometric masses of Super-Schmidt meteors; this study is still the source of luminous efficiencies commonly in use \citep[e.g.][]{Ceplecha1976}. \cite{Verniani1965} assumed that luminous efficiency is proportional to meteoroid speed to some constant exponent ($\tau \propto v^n$), which his data found to be $n = 1$. \cite{Verniani1965} made an effort to correct his results for fragmentation as it was well known that many of the Super-Schmidt meteors crumbled during ablation, but his calibration for $\tau$ is based on a single non-fragmenting asteroidal meteor at low altitude. Lab experiments entail charging and accelerating tiny metal particles in a Van deGraaf generator and observing as they ablate in a low-pressure air chamber. A common difficulty in completing these lab studies is accurately recreating atmospheric compositions and conditions. A number of studies were carried out in the late sixties and early seventies \citep[see][]{Friichtenicht1968,Becker1971,Becker1973}, and work in this area has recently been revived, though no results for luminous efficiency have been published yet \citep{Thomas2016}. Another method of determining luminous efficiency is to use artificial meteoroids, in which objects of known mass, composition, and density are launched into the atmosphere and observed as they ablate. This was done by \cite{Ayers1970} for iron and nickel projectiles, at relatively low speeds compared to meteors. Simultaneous radar and optical studies, which use the ionisation efficiency to determine the luminous efficiency, were carried out by \cite{Saidov1989} (who assumed the optical data was more correct) and \cite{Weryk2013} (who assumed the radar data was more correct), but the results of the two studies do not agree. A combination of theoretical and lab work was analysed by \cite{Hill2005}, and the result (corrected for bandpass and composition) is presented in \cite{Weryk2013}. \newline

Many of these studies are summarised in \cite{Subasinghe2017}, who investigated the precision of using the dynamic mass to determine the photometric mass using meteor data simulated with the ablation model of \cite{CampbellBrown2004}. Data with sub-metre scale resolution from the Canadian Automated Meteor Observatory (CAMO) has the potential to redo the work of Verniani much more accurately. In testing the method, we found that uncertainties in the meteoroid density (which cannot be uniquely determined from observations) and atmospheric density account for a factor of two uncertainty, each, in the calculated luminous efficiency. A similar factor of two uncertainty was found when investigating simulated meteors of different masses and speeds. Good agreement was found between the luminous efficiency value used in the simulations and the derived values for all of the mass-speed groups except for the lowest speed group for each of the three masses, which had slightly poorer agreement. \cite{Subasinghe2017} used simulated data free from measurement noise, so the effects of observed noise were not considered, or the possibility that parameters in Equation \ref{eq:drag} may not be constant over time. \newline

In this work, we continue that investigation by applying a modified method to real meteor events recorded with the high-precision Canadian Automated Meteor Observatory.

\section{Method Refinement}

The simulated data used in \cite{Subasinghe2017} was free of noise, which is unrealistic, but made it easy to evaluate various functional fits to the meteor lag, or distance by which the meteor lags behind an object travelling at constant speed. That study found that a two-term exponential fit to the lag matched the luminous efficiency most closely. The real data used in this study have residuals in position of about 2 m: we added scatter of this order to the positions generated by the ablation model, and again tried to recover the luminous efficiency used in the model. A simplification from the two-term exponential (lag = $a\exp{bx} + c\exp{dx}$) to a single-term exponential (lag = $a\exp{bx}$) was necessary, because when noise was added to the model the two-term exponential produced unphysical values of the speed and deceleration. A single-term exponential is a relatively poor fit for the full lag curve, but does well when fitting only the last half of the curve, which is where the deceleration is high enough to calculate a meaningful luminous efficiency. The final method used in this work fits a single-term exponential to the second half of the meteor lag data, and uses the fit parameters to determine the speed and deceleration profiles used for the luminous efficiency determination. \newline

To illustrate the process, we present a standard, non-fragmenting simulated event (with the parameters given in Table \ref{tab:standard}) run through our method, and the resulting luminous efficiency profile. The event was simulated using the meteoroid ablation model of \cite{CampbellBrown2004} with an assumed luminous efficiency of 0.7\%, constant over the entire ablation time. The derived luminous efficiency was calculated using Equations \ref{eq:luminous} and \ref{eq:drag}, with values for the drag coefficient $\Gamma$, meteoroid density $\rho_{m}$, shape factor $A$, and atmospheric density $\rho_{atm}$ coming from the simulation (for real events, these will be estimates except for the atmospheric density which will come from a model). A single-term exponential (given in Equation \ref{eq:lag}) was fit to the second half of the lag data of the standard event, and the fitted parameters were $a = 1.112$ and $b = 6.84$. 

\begin{equation}
\label{eq:lag}
\textrm{lag} = a\exp(bx)
\end{equation}

\begin{figure}
\epsscale{0.75}
\plotone{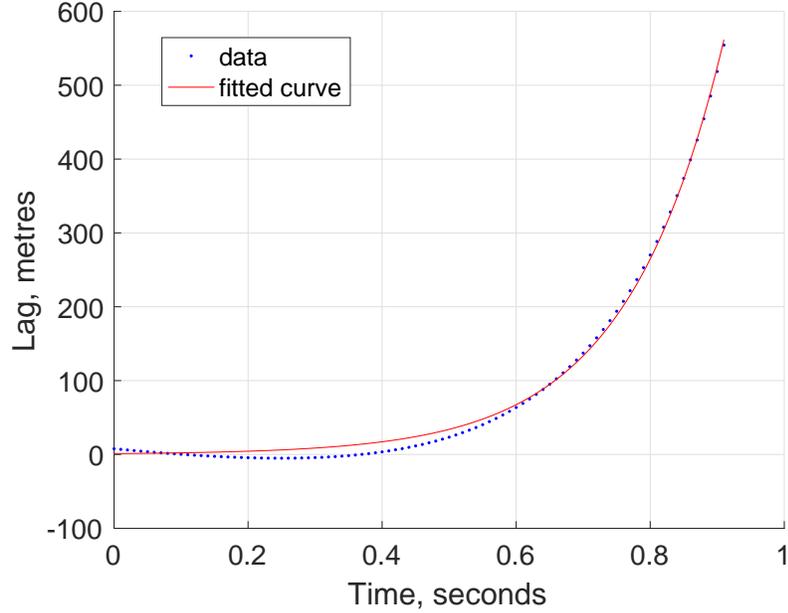}
\caption{The second half of the lag (from 0.45 seconds onward) was fit with a single-term exponential. The fit has been plotted over the entire lag profile (shown as blue dots). The negative lag points are due to the method used to determine the lag -- if we fit fewer than 50\% of the meteor position points to determine the initial speed, there would be no negative lag values.}
\label{fig:lag_fit}
\end{figure}

\begin{figure}
\plottwo{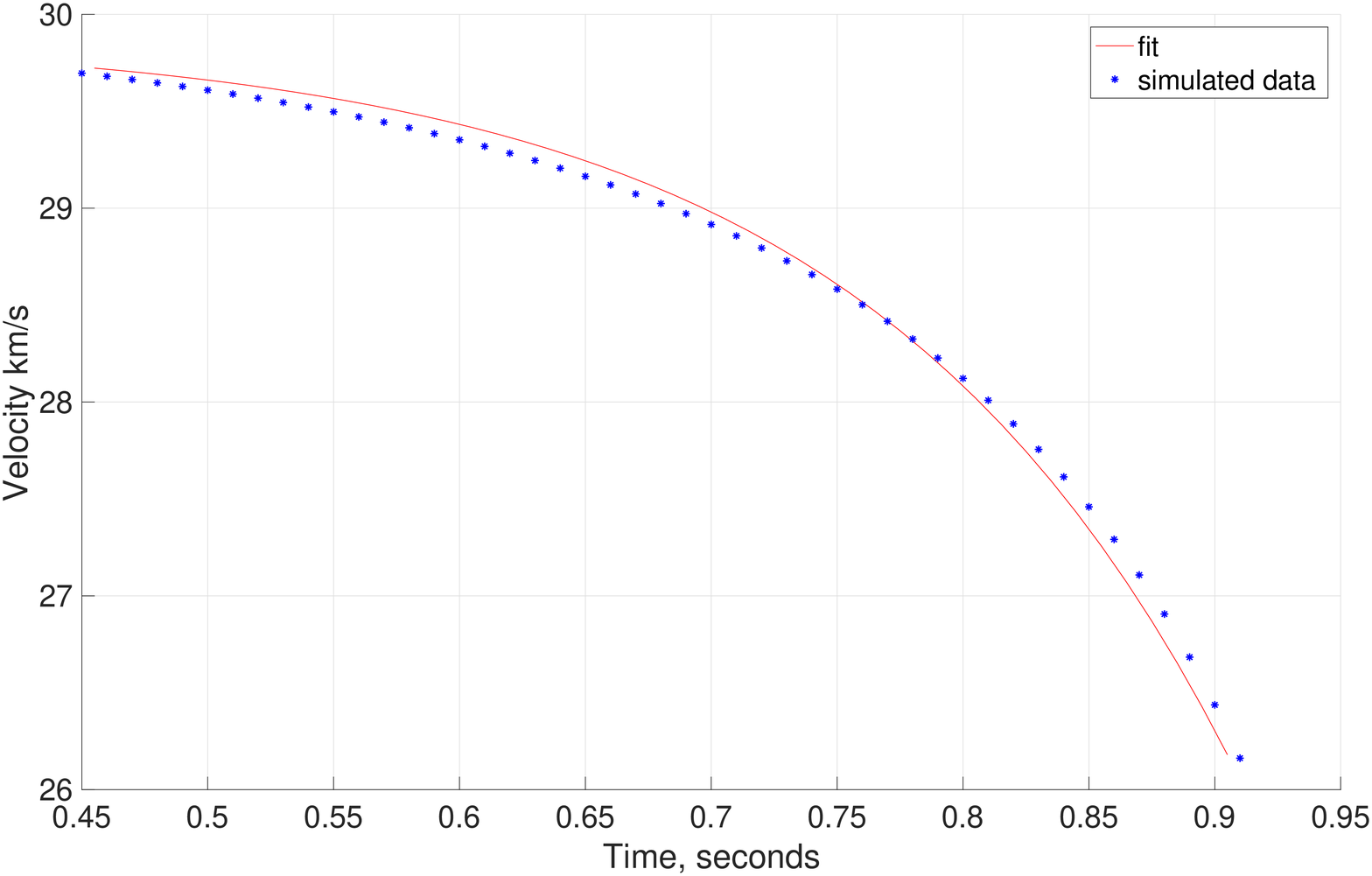}{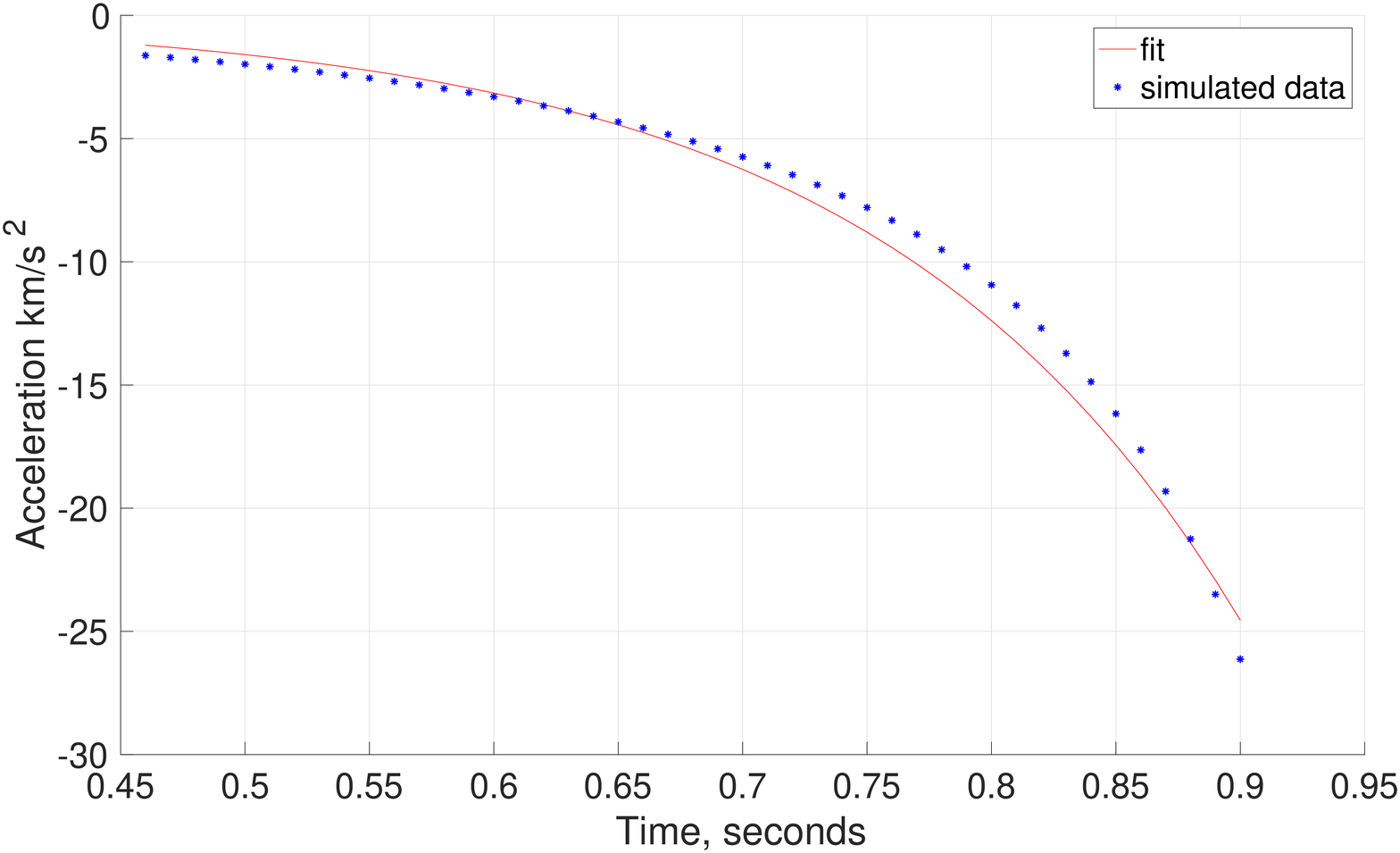}
\caption{The speed profile of the standard simulated event, with the simulated data shown as blue asterisks, and the velocity based on the lag fit shown as a solid red line is on the left. On the right is the deceleration profile for both the simulated data and the data based on the lag fit, shown with blue asterisks and a solid red line respectively. In both plots, only the second half of the simulated meteor data is shown, as that was all that was fit by the single-term exponential.\label{fig:vel_acc}}
\end{figure}

The speed and deceleration values are based on these fitted parameters. The fit to the lag is shown in Figure \ref{fig:lag_fit}, and the corresponding speed and deceleration plots are shown in Figure \ref{fig:vel_acc}, with the simulated values plotted against the curves based on the fit parameters from the lag. \newline

\begin{deluxetable}{cccccc}
\centering
\tablecaption{Standard event parameters.\label{tab:standard}}
\tablehead{ \colhead{Initial Speed} & \colhead{Initial Mass} & \colhead{Shape Factor} & \colhead{Drag Coefficient} & \colhead{Meteoroid Density} & \colhead{Zenith Angle}\\ 
\colhead{(kms$^{-1}$)} & \colhead{(kg)} & \colhead{} & \colhead{} & \colhead{(kgm$^{-3}$)} & \colhead{(degrees)}} 
\startdata
30 & 10$^{-5}$ & 1.21 & 1 & 2000 & 30 \\
\enddata
\tablecomments{These are the parameters used in the \cite{CampbellBrown2004} ablation model to simulate our standard event for testing purposes.}
\end{deluxetable}

\begin{figure}
\epsscale{0.75}
\plotone{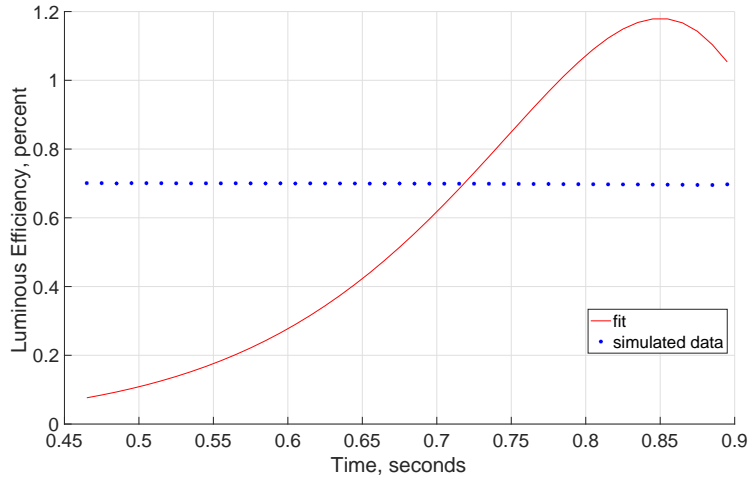}
\caption{The luminous efficiency profile for the simulated meteor based on the simulated speed and deceleration profiles is shown with blue asterisks, and the luminous efficiency profile based on the speed and deceleration derived from the fitted lag is shown with a solid red line. To quantify the luminous efficiency for each profile, the mean value was used, based on the second half of the meteoroid ablation data.}
\label{fig:tau}
\end{figure}

Using the speed and deceleration profiles based on the lag fit, and all other values taken from the simulation, we find the luminous efficiency profile shown in Figure \ref{fig:tau}. Recall that in the simulation, a constant luminous efficiency of 0.7\% was used. Figure \ref{fig:tau} shows two profiles: the blue asterisks show the luminous efficiency determined using all the simulation parameters, including the simulated deceleration and simulated speed. The red line shows the luminous efficiency found using the speed and deceleration profiles based on the single-term exponential fit to the lag. The resulting profile from the simulated data is not able to perfectly reproduce the constant 0.7\% luminous efficiency used in the simulation due to the simplicity of this method compared to the physics in the ablation model that was used. The average luminous efficiency of the profile based on the fitted lag is 0.598\%, and 0.699\% for the profile resulting from using all the simulated data. \cite{Subasinghe2017} explored the deviation from an assumed luminous efficiency due to different functional fits and each parameter in the drag and luminous intensity equations. 

\section{Equipment and Data Reduction}
\subsection{Equipment}
The meteor events in this study were recorded by the Canadian Automated Meteor Observatory (CAMO), which is located in Ontario, Canada. CAMO is a two-station image intensified optical system which records meteors nightly, under appropriate weather conditions. One station is located in Tavistock, Ontario, Canada (43.265$^\circ$N, 80.772$^\circ$W), and the other is about 45 km away in Elginfield, Ontario, Canada (43.193$^\circ$N, 81.316$^\circ$W). For our work we are using the guided system which is a two-part system, consisting of a wide-field camera (with a field of view of 28$^\circ$) and a narrow-field camera (1.5$^\circ$ FOV). The wide-field camera detects meteors in real time with the All Sky and Guided Automatic Realtime Detection (ASGARD) software \citep{Weryk2008} which guides a pair of mirrors to direct the meteor light into the narrow-field camera. The cameras run at different frame rates (80 and 110 fps for the wide-field and narrow-field, respectively), and each camera is used for different science goals. The wide-field camera typically captures the full meteoroid ablation profile, which allows for light curve calculation, as well as orbit determination. The narrow-field camera does not capture the entire meteoroid ablation profile as collection only begins after the meteor has been detected in the wide-field camera over 4 to 7 frames, and it typically takes a few frames of observation in the narrow-field camera before the meteor is tracked smoothly by the mirrors. However, the narrow-field camera is a high-resolution camera able to resolve up to 3 m per pixel at 100 km range, which allows the fragmentation behaviour of the object to be observed. These observations allow us to find objects that show single-body ablation, which is required when using the classical ablation equations, and to measure the deceleration very precisely. More details about the cameras can be found in \cite{Weryk2013a}.

\subsection{Data Reduction}
\begin{figure}
\epsscale{0.3}
\plottwo{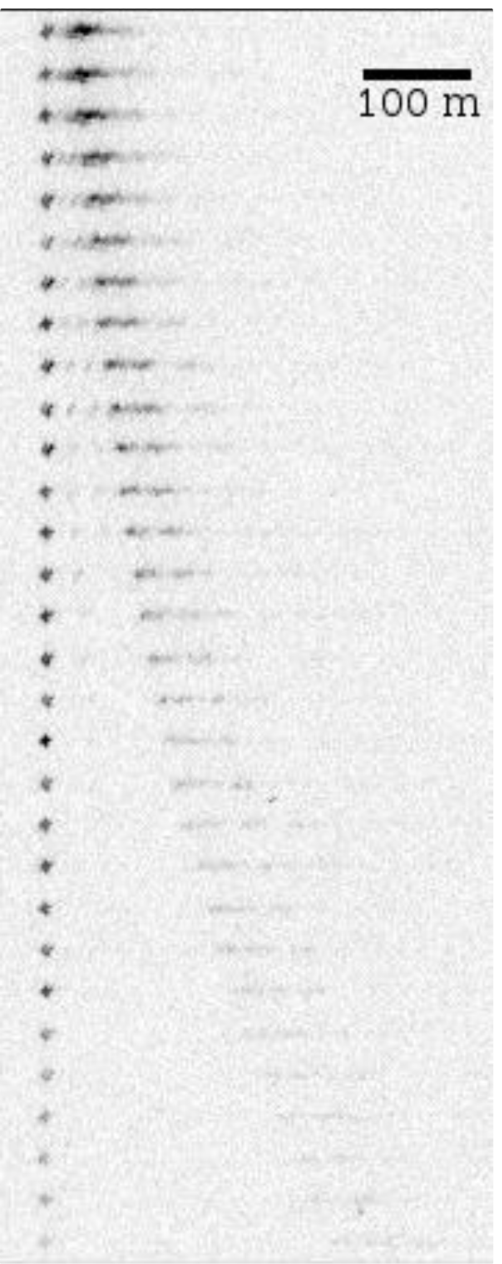}{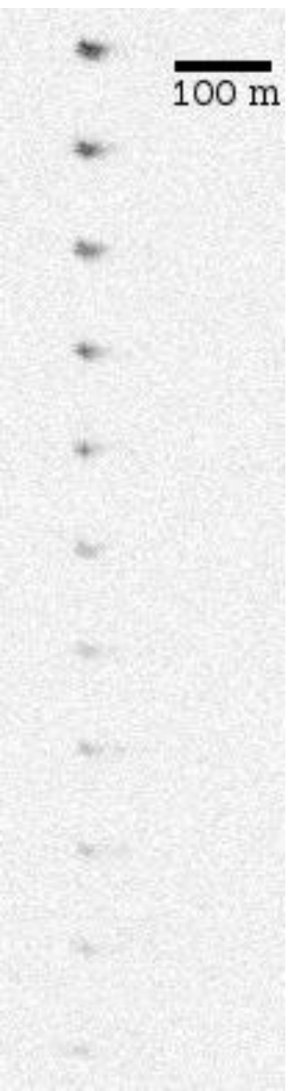}
\caption{Examples of a meteor with a leading fragment and another showing single body ablation. Each image is from an individual frame, with the meteor cropped, rotated, and stacked such that time increases downwards. The image on the left has the leading fragment centered (clearly showing the deceleration of the rest of the body behind it), while the meteor on the right has the entire object centered. The 100 m scale bar for the leading fragment example (on the left) is for a height of 80 km, and for the single body example on the right, is for a height of 85 km. }
\label{fig:morphology}
\end{figure}

CAMO records meteors each night if certain sky conditions are met, and has been in operation since 2007 \citep{Weryk2013a}. This large database of meteors was searched for meteor events recorded at both stations in the narrow-field cameras that showed either single body ablation, or a leading fragment with no wake. The classical meteoroid ablation equations apply to objects that are solid, single bodies that do not fragment. Any object which does fragment will have a smaller dynamic mass (this is the mass of the largest and brightest fragment) than the photometric mass (this is the mass of the entire meteoroid), and therefore an artificially large luminous efficiency. Despite CAMO having recorded thousands of meteor events, finding meteors that showed next to no fragmentation proved to be a difficult task, as more than 90\% of CAMO meteors show some form of fragmentation in the form of wake \citep{Subasinghe2016}. A meteor with a leading fragment is a special case where a fragment of the main body decelerates less than the others, and this piece shows little to no fragmentation. Examples of these non-fragmenting morphologies are shown in Figure \ref{fig:morphology}.  Meteors that met the requirements of showing single-body ablation (either the entire body, or a leading fragment) were then analysed with two software packages: METAL \citep{Weryk2012} and mirfit. \newline

METAL allows for orbit determination and light curve analysis of a meteor in the wide-field camera. Astrometric and photometric plates are computed using a minimum of 10 stars from each station: each stellar pixel centroid and brightness are calibrated against those from the SKY2000v4 catalogue. Once this is complete, the head of the meteor is picked in each frame, for both stations, for which the entire meteor is visible, and a trajectory solution is determined using an implementation of the least squares method \citep{Borovicka1990} called MILIG. By masking out pixels containing light from the meteor, the meteor apparent magnitude can be determined, and converted to an absolute magnitude using the photometric calibration plate and meteor range. \newline

The software package mirfit allows video observations taken with the high-resolution narrow-field cameras to be analysed. Because the field of view is so small (1.5$^\circ$) compared to video taken with the wide-field cameras (28$^\circ$), stellar astrometric plates cannot be done due to the lack of visible stars; in addition, the field of view is moving during the observations. To determine meteor positions in the narrow-field cameras, the mirror positions need to be taken into account. These are recorded every 0.5 ms, so the position of the centre of the field at the beginning of the exposure can be interpolated. Then, the distance (or offset) between the position of the meteor on the narrow-field image and the image centre is determined, and converted into an offset in mirror encoder coordinates. This offset is then added to the mirror position at that time, and mapped onto the celestial sphere. This is done using two plates: a scale plate which determines the offset and converts pixel position to mirror encoder coordinates; and an exact plate which maps the mirror encoder coordinates into celestial coordinates. A calibration for the exact plate is done at the beginning of any night's observations, and every two hours through the night.\newline

If the program that creates the exact plate calibration makes an error, for example by attempting to calibrate a star in the field of view with a different star in the catalogue, errors with the plates can occur, so plates are verified prior to meteor analysis, since bad calibration data cannot be replaced. Stars visible in the field of view will be trailed across each frame, with the predicted location of the initial position indicated by the software. If the plates are functioning correctly, the predicted star positions will not drift across the star trails, but will have the same position relative to the star streak in each frame. The method we are using to determine meteor luminous efficiency is very sensitive to the position measurements (as deceleration values are needed), which means for the most reliable results, we should only use meteor events with accurate plates. The number of useful meteors is then restricted not only to the few meteors which show no fragmentation, but to those non-fragmenting meteors which also have accurate plates. Meteor events that pass both of these requirements are then analysed in mirfit. The meteor astrometry is done through centroiding for each frame, in both stations; centroiding works particularly well on meteors with no visible wake. Photometry was also done for each meteor. mirfit is not able to calibrate the meteor brightness because of the lack of calibration stars in the field, but can calculate the log of the sum of the brightness of the meteor pixels, which is proportional to the meteor magnitude. This relative light curve is then calibrated using the photometry for the meteor in the wide-field camera. \newline

\begin{figure}
    \centering
    \includegraphics[scale = 0.3,angle = 90]{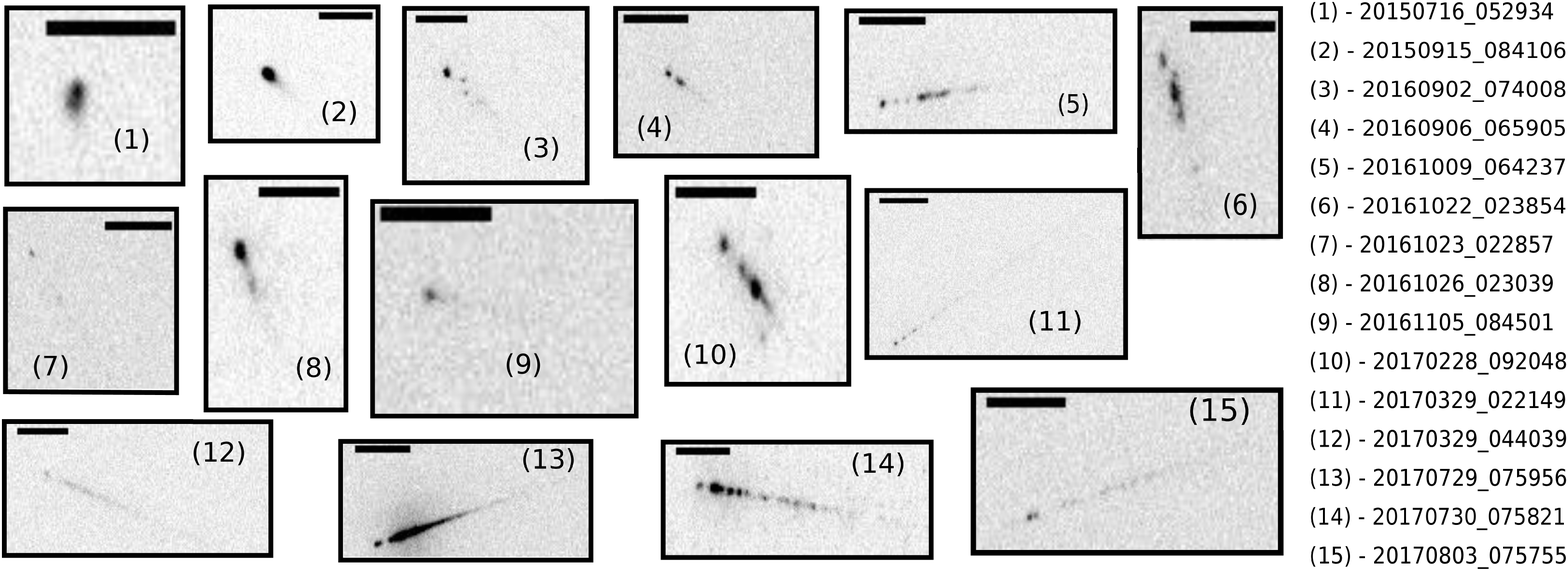}
    \caption{A single frame from the analysed meteor events. The scale bar in each frame corresponds to 100 m. In frames showing multiple fragments, the fragment of interest is the front most one. Images have been inverted to show detail.}
    \label{fig:frames}
\end{figure}

Data reduction was completed for thirteen meteors showing a leading fragment, and two meteors showing single body ablation. A single frame from each meteor is shown in Figure \ref{fig:frames}. Five fragmenting meteors were also analysed for comparison, though the luminous efficiencies calculated for those meteors must be upper limits: the dynamic masses will be less than the photometric masses, leading to larger derived luminous efficiency values. The mirfit analysis provides high-precision meteor positions, with an average random position uncertainty of 1.6 m for the fifteen non-fragmenting events (with a maximum uncertainty of 2.5 m, and a minimum of 0.9 m). The position data was turned into lag values (i.e. the distance the meteoroid would fall behind an identical object moving with a constant speed). The speed used was the initial meteoroid speed, found by fitting the first half of the distance-time data, though the exact value is not important since only derivatives of the lag are used. The second half of this lag was fit by a single-term exponential function, and the derivatives were used for the speed and deceleration profiles. Typical values were used for the drag coefficient ($\Gamma$ = 1); shape factor ($A$ = 1.21); meteoroid density (values taken from either \cite{Kikwaya2011a} or \cite{Kikwaya2011}); and the atmospheric density profile was taken for the event date from the NRLMSISE-00 Atmosphere model \citep{Picone2002}. An analysis of how sensitive this method is to each of the parameters can be found in \cite{Subasinghe2017}.

\section{Results}
\subsection{Noise analysis}
\begin{figure*}
\gridline{\fig{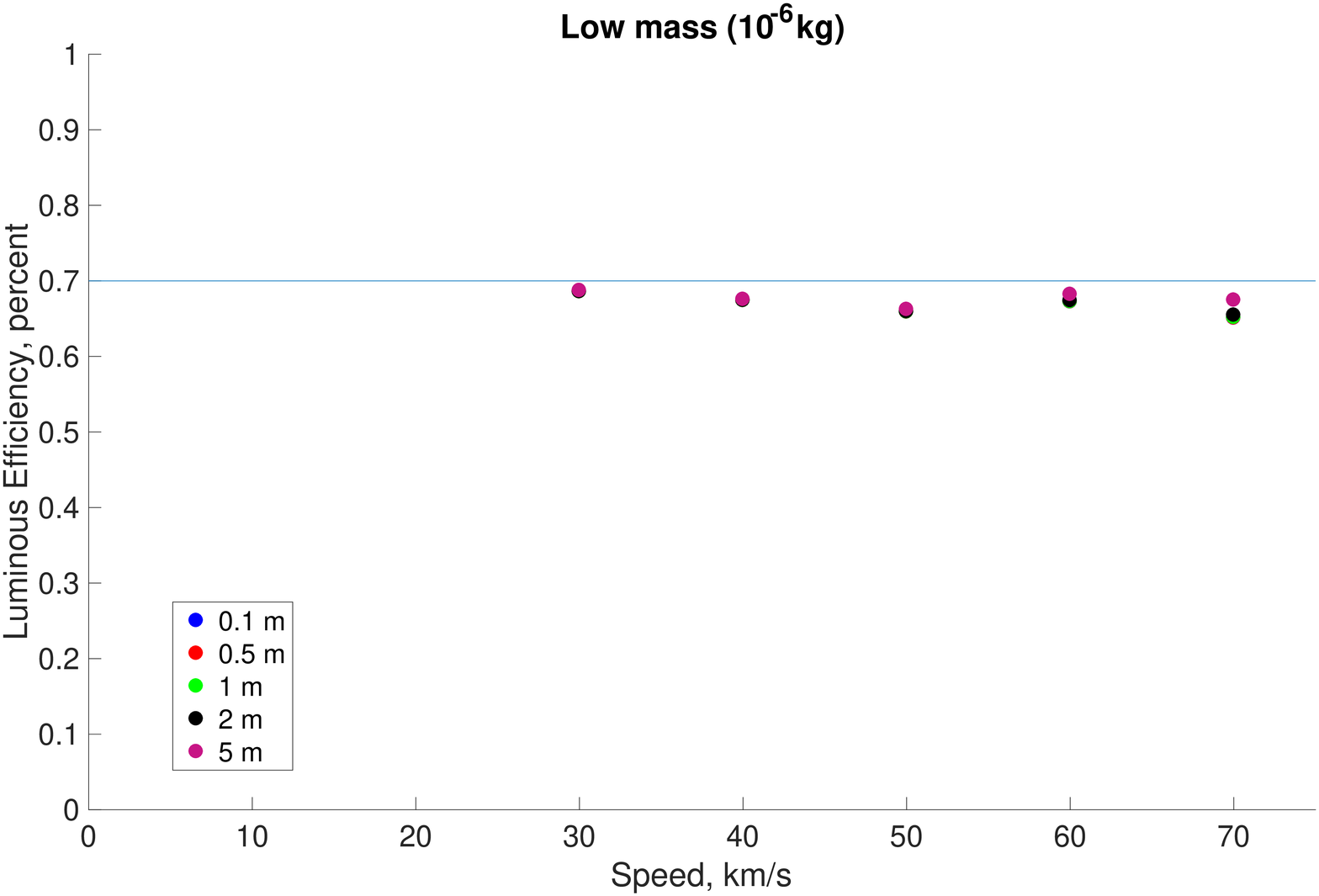}{0.4\textwidth}{(a)}
\fig{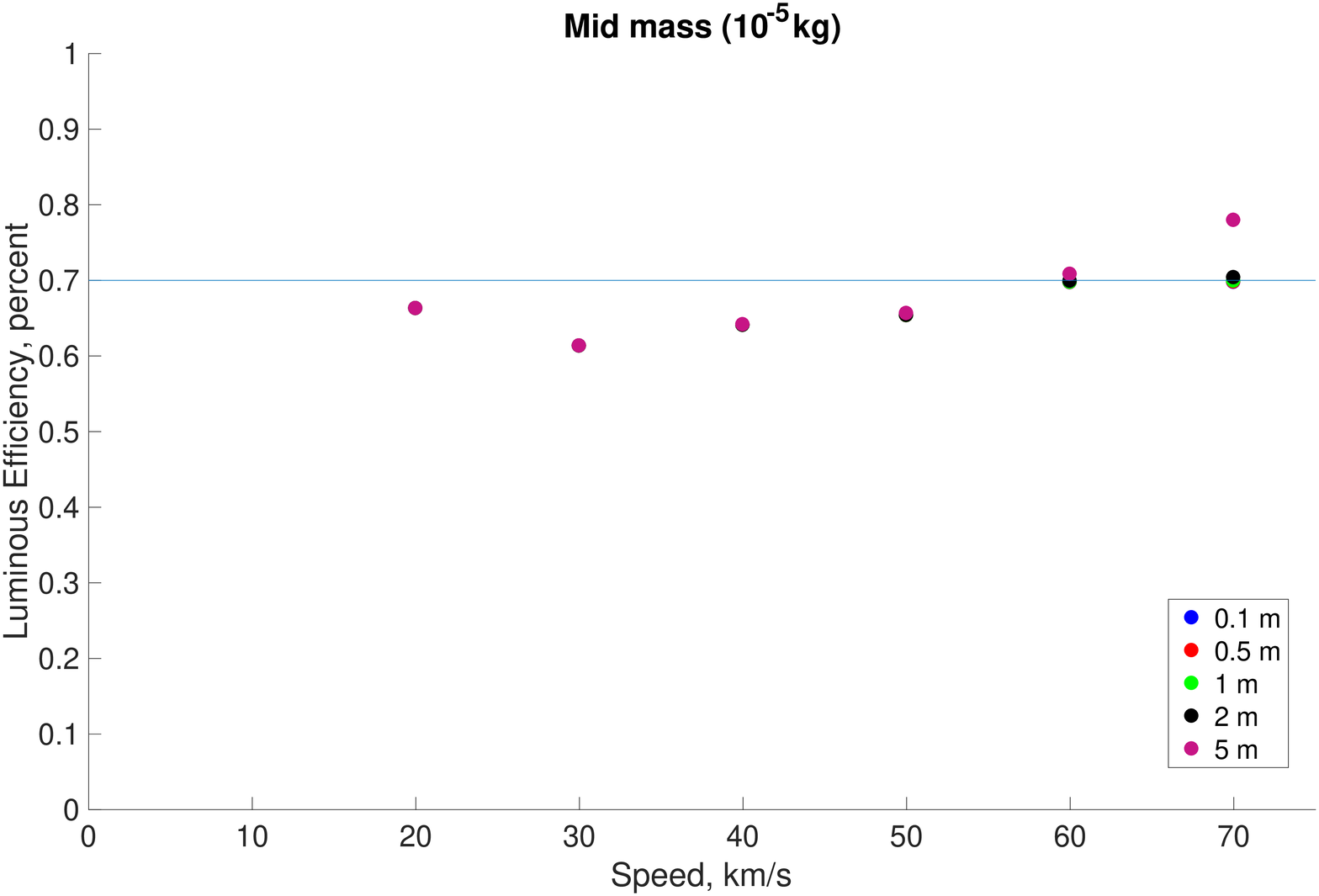}{0.4\textwidth}{(b)}}

\gridline{\fig{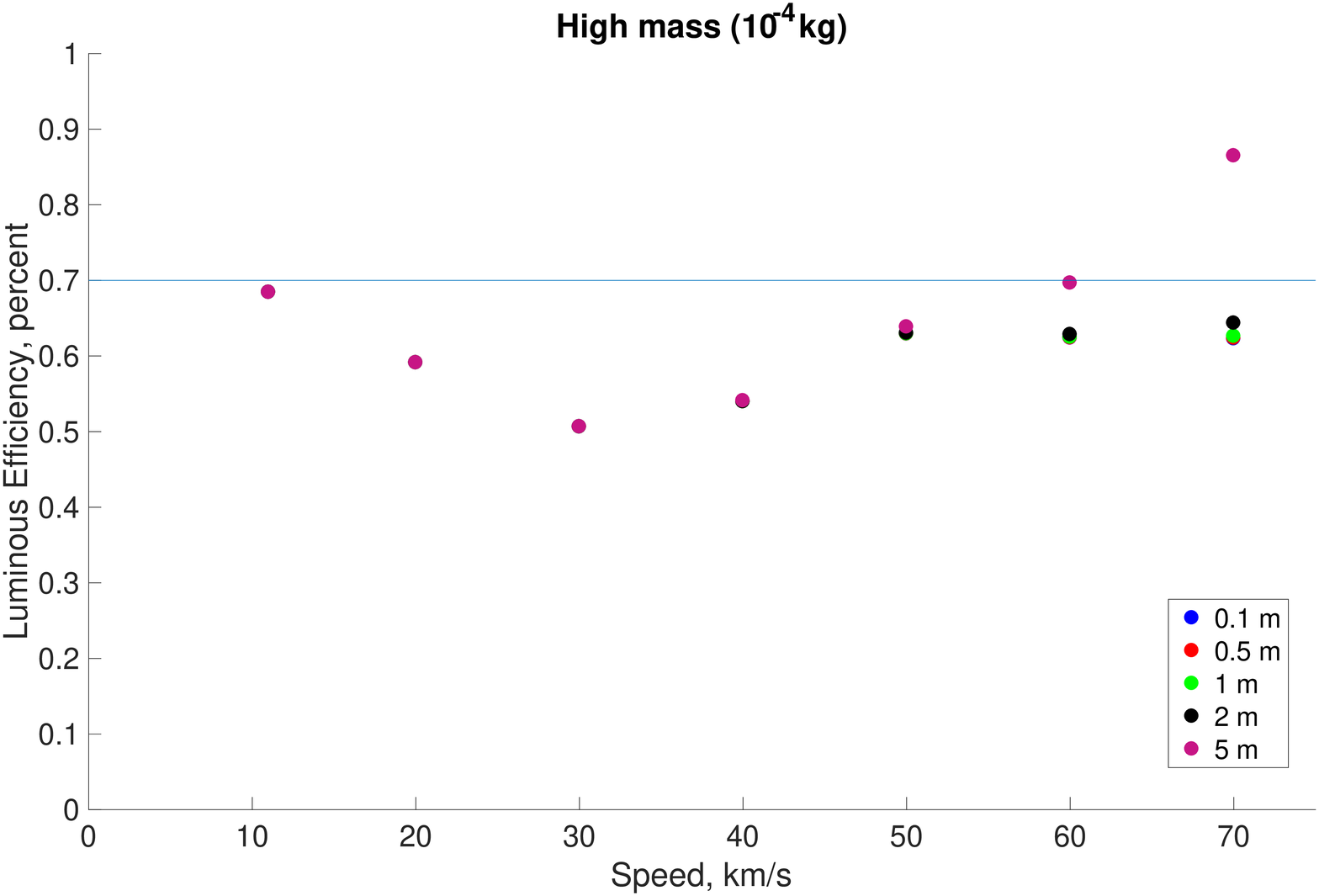}{0.4\textwidth}{(c)}}
\caption{Different amounts of uncertainty were added to simulated meteor position picks, and the resulting derived averaged luminous efficiencies are shown for three different mass groups. Each meteor was simulated with a constant luminous efficiency of 0.7\%, shown as a solid blue line. The different coloured points indicate the position uncertainty. \label{fig:noise}}
\end{figure*}

Prior to evaluating the luminous efficiency for each of our fifteen events, we investigated the effect that noise has on our method. In \cite{Subasinghe2017} we investigated a set of simulated meteor events that covered the entire physical phase space of mass, speed, meteoroid density, zenith angle, and shape factor. There were 21 mass-speed groups (three different masses and seven different meteor speeds), each of which had 50 possible meteors (all combinations of five possible meteoroid densities, two possible zenith angles, and five possible shape factor values; the possible values are given in Table \ref{tab:simulated}); however, not all meteors produced enough light that the CAMO optical system would detect it. This left 18 mass-speed groups, with up to 50 meteors, to study. Each meteor was simulated with a luminous efficiency of 0.7\%, constant over time. Five uncertainty levels (standard deviations of 0.1, 0.5, 1, 2, 5 m) were selected to showcase the effect that position precision would have on the results, with the 2 m uncertainty being closest to our measured uncertainties. Noise was randomly added 500 times at each uncertainty level to each meteor in the 18 mass-speed groups, and the luminous efficiency was calculated. The mean value of each profile was found, and the 500 mean luminous efficiency values for each event were averaged. These results are presented in Figure \ref{fig:noise}, separated by mass.

\begin{deluxetable}{cccccc}

\tablecaption{Parameters used to simulate meteors.\label{tab:simulated}}
\tablehead{\colhead{Mass} & \colhead{Speed} & \colhead{Meteoroid density} & \colhead{Zenith Angle} & \colhead{Shape factor} & \colhead{} \\ 
\colhead{(kg)} & \colhead{(kms$^{-1}$)} & \colhead{(kgm$^{-3}$)} & \colhead{(degrees)} & \colhead{} & \colhead{} } 

\startdata
10$^{-4}$ & 11 & 1000    & 30      & 0.5 \\ 
10$^{-5}$ & 20 & 2000    & 60      & 0.8 \\ 
10$^{-6}$ & 30 & 3000    & \nodata & 1.21 \\ 
\nodata   & 40 & 5000    & \nodata & 2 \\ 
\nodata   & 50 & 8000    & \nodata & 4 \\
\nodata   & 60 & \nodata & \nodata & \nodata\\ 
\nodata   & 70 & \nodata & \nodata & \nodata\\
\enddata
\end{deluxetable}

\subsection{Atmospheric Density variations}
\begin{figure}
\plotone{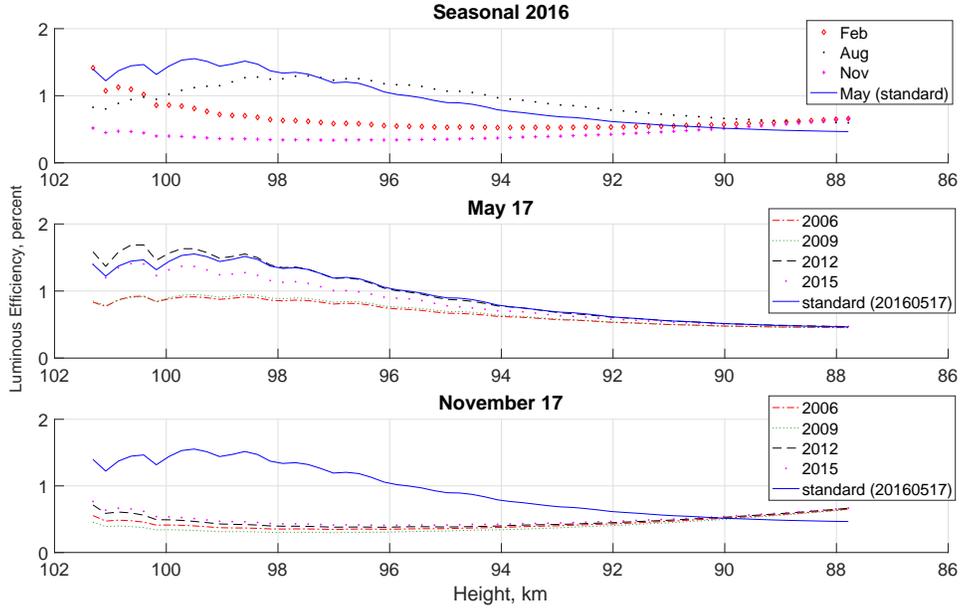}
\caption{Atmospheric density variations on a simulated meteor event. The top panel shows seasonal changes over 2016. The middle and bottom panels show solar cycle changes. All values for the derived luminous efficiency profiles came from the simulation, except for the seasonal and yearly atmospheric density models. There are minimal differences in derived luminous efficiency profiles over the solar cycle, which may be due to a lack of data at meteoroid ablation heights.}
\label{fig:atm_variations}
\end{figure}

As an extension to the work presented in \cite{Subasinghe2017}, we investigated the influence of atmospheric density changes on derived luminous efficiency values. It was found that changes in the model atmospheric density profiles over the course of one year could affect derived luminous efficiency values by at most a factor of two (with the model atmospheric density profiles varying at different heights by at most a factor of 2). Figure \ref{fig:atm_variations} shows in the top panel how seasonal variations over the course of 2016 would affect the derived luminous efficiency profile of our standard event. Everything was identical in each run except for the atmospheric profile used; in the standard profile for comparison, the atmospheric density profile used in the simulation was used to find luminous efficiency. All other values (meteoroid density and so on) matched those used in the simulation. The middle and bottom panels instead show how the solar cycle affects the luminous efficiency for two dates over the course of 2006 - 2012. At the time of writing, the NRLMSISE-00 model provides data up to April 17 2017. However, some of our meteor events were recorded in July and August 2017. The results of Figure \ref{fig:atm_variations} indicate that atmospheric density data from the same day but different years will result in very similar derived luminous efficiency results: the May 17 derived mean luminous efficiency results vary at most by a factor of 1.4, and the Nov 17 derived mean luminous efficiency values vary at most by a factor of 1.3. Therefore, for our meteor events recorded in July and August 2017, atmospheric density profiles from the previous year were used. 

\subsection{Photometry calibration}
When determining the luminous efficiency with the drag and luminous intensity equations, the meteoroid brightness is needed, as seen in Equation \ref{eq:luminous}. This information can be obtained from both the narrow-field and wide-field cameras; however, if the fragment of interest is a leading fragment, the wide-field photometry will include the brightness of all fragments (the resolution is not high enough to separate the fragment brightness from the rest of the object). Thus, the narrow-field photometry is necessary for determining the luminous efficiency of the relevant fragment only. The method of obtaining the meteor photometry from both the wide-field and narrow-field cameras is described below.

The calibration of meteor photometry for meteors observed with the CAMO guided wide-field system was discussed in \cite{Weryk2013a}. As we are using positions derived from the narrow-field analysis, we investigated the possibility of calibrating the narrow-field meteor photometry with stars observed in the narrow-field camera, to eliminate the need for the wide-field cameras in this work. In METAL, pixels are masked out in each frame to select the light from the meteor (giving the instrumental apparent magnitude), and converted to the absolute magnitude using both the previously determined photometric plate and range to the meteor in each frame. The uncertainty in METAL photometry is close to 0.2 mag \citep{Subasinghe2016}. Similarly, in mirfit, pixels can be masked out in each frame to give the instrumental apparent magnitude from the log of the sum of the pixel values ($lsp$); however, there is no photometric plate due to the small number of visible stars in the small field of view. \newline

To calibrate the narrow-field instrumental apparent magnitudes, we compared those log-sum-pixel values to the absolute magnitudes determined in the wide-field observations and used them as a calibration to determine the absolute magnitude of the meteor in the narrow-field camera. To verify these narrow-field absolute magnitudes, we investigated the method of calibrating against stars visible in the narrow-field camera, in spite of their small number. \newline

Photometry calibrations were investigated for seven of the meteor events in our dataset (four events had no visible stars, and nine events were added after this investigation was completed). Each investigated event had at least one visible star, but no more than two - more than two visible stars occurred multiple times, but stars were eliminated from the study if they were only visible for a few frames, or if they were binary stars. The average instrumental apparent magnitude of each star was determined, and the offset from the SKY2000 catalogue R magnitude was found. This offset was applied to the meteor $lsp$ values to correct them to apparent magnitude values. If a station had two visible stars the average offset was applied to the meteor log sum pixel values to correct them to an apparent magnitude. The average difference between the mirfit calibrated photometry and the METAL calibrated photometry was -0.3098 and -0.2666 magnitude, for Tavistock and Elginfield respectively. For our luminous efficiency analysis, we used the METAL calibrations simply because there are many more stars to calibrate against, but our result here indicates that there is only a minor difference between the METAL and mirfit brightness calibrations which are typically based on two stars.

Equation \ref{eq:luminous} uses the meteor intensity, rather than the meteor magnitude. To convert between magnitude and intensity, we use the results of \cite{Weryk2013} who determined that for the Gen III image intensified video cameras we are using, a zero magnitude meteor emits 820 W.

\subsection{Meteoroid density} \label{meteoroid_density}
One parameter we have control over in our analysis is the initial meteoroid density. In our study we assume for simplicity that this density is constant over time. The density for each meteoroid was determined using the results from \cite{Kikwaya2011a} and \cite{Kikwaya2011}, in which meteoroid densities were found after searching the entire parameter space in an attempt to match each meteor deceleration and light curve shape using the ablation model of \cite{CampbellBrown2004}. \cite{Kikwaya2011a} used the classification of \cite{Borovicka2005}, which considers meteoroid physical composition, in their meteoroid density analysis, and uses not only the meteoroid Tisserand parameter, but individual orbital elements to help classify objects. The Tisserand parameter with respect to Jupiter is based on an object's orbital elements, and is given by:

\begin{equation}
T_J = \frac{a_J}{a} + 2\sqrt{\frac{a}{a_J} (1 - e^2)} \cos(i)
\end{equation}

Where $a$, $e$, and $i$ are the semimajor axis, eccentricity, and inclination of the meteoroid, respectively, and a subscript $J$ indicates an orbital element belonging to Jupiter. A Tisserand parameter greater than 3 suggests that an object has an asteroidal orbit; between 2 and 3 is associated with Jupiter-family comets; and $T_J$ less than 2 describes a Halley-type orbit. \newline

The meteoroid orbit classification described by \cite{Borovicka2005} and the associated meteoroid densities from \cite{Kikwaya2011a} used in this work is given in Table \ref{tab:tisserand} (note: $q$ is the perihelion distance and $Q$ is the aphelion distance). \newline

In our set of non-fragmenting meteor events we have eight meteors with Halley type orbits, five with Jupiter family orbits, and two with asteroidal-chondritic orbits. Of our five fragmenting meteor events, two are Halley-type, one has a Jupiter family orbit, and two have asteroidal-chondritic orbits. \cite{Kikwaya2011a} give a mean density for objects with Jupiter family orbits and objects with asteroidal-chondritic orbits. They give a minimum and maximum density for objects on Halley type orbits: the mean density for Halley type objects comes from the raw data from \cite{Kikwaya2011}. 

\begin{deluxetable}{ccc}
\tablecaption{Meteoroid densities based on Tisserand parameter values. \label{tab:tisserand}}
\tablehead{\colhead{Orbit type} & \colhead{Orbital element and Tisserand parameter} & \colhead{Density} \\ 
\colhead{} & \colhead{} & \colhead{(kgm$^{-3}$)} } 
\startdata
Sun-approaching & q $<$ 0.2 AU & 3206 \\
Ecliptic shower  & e.g. Northern Iota Aquariids& 3200\\
Halley type  & $T_J$ $<$ 2 or 2 $<$ $T_J$ $<$ 3 and $i$ $>$ 45$^\circ$  & 890\\
Jupiter family & 2 $<$ $T_J$ $<$ 3 and $i$ $<$ 45$^\circ$ and $Q$ $>$ 4.5 AU  &3190 \\
Asteroidal-chondritic  & $T_J$ $>$ 3 or $Q$ $<$ 4.5 AU  & 4200\\
\enddata

\tablecomments{The first two columns are from \cite{Borovicka2005}, and the associated meteoroid densities are based on work from \cite{Kikwaya2011} and \cite{Kikwaya2011a}.}
\end{deluxetable}

\subsection{Error analysis}

In Table \ref{tab:results} we present the luminous efficiency values determined for each of the twenty events. Each luminous efficiency value is presented with an associated uncertainty which takes into account the uncertainty from assuming the drag coefficient, meteoroid density, shape factor, and random errors in the position. A random error of up to half a pixel was added to each analysed position pick, one hundred times. Half a pixel corresponds to approximately 2 m at a range of 110 km, which was found to be closest to the average error in our measured positions. The entire parameter space of drag coefficient, meteoroid density, and shape factor was then tested with each of those one hundred variations, and the luminous efficiency was found. Table \ref{tab:results} presents the mean luminous efficiency and corresponding standard deviation. The range of drag coefficient values tested was from 0.8 to 1.2 (with 1.0 being an inelastic collision); shape factor varied from 0.71 to 1.71 (with 1.21 being a sphere); and meteoroid density taken from \cite{Kikwaya2011a} with an uncertainty of $\pm$ 500 kgm$^{-3}$. The distribution of luminous efficiency values spanning the entire parameter space was more skewed than normal, however Table \ref{tab:results} presents the mean, median, and standard deviation values.

\begin{deluxetable}{cccccccccc}

\tablecaption{Parameters for the 20 meteors. \label{tab:results}}
\tabletypesize{\scriptsize}
\tablehead{\colhead{Event} & \colhead{v$_i$} & \colhead{T$_J$} & \colhead{Orbit type} & \colhead{Origin} & \colhead{$\rho_m$} & \colhead{m$_i$} & \colhead{$\tau$ (mean)} & \colhead{$\tau$ (median)} &\colhead{Note} \\ 
\colhead{} & \colhead{(kms$^{-1}$)} & \colhead{} & \colhead{} & \colhead{} & \colhead{(kgm$^{-3}$)} & \colhead{(10$^{-6}$ kg)} & \colhead{(per cent)} & \colhead{(per cent)} & \colhead{} } 

\startdata
20160902\_074008 & 68.9 & -0.2 & HT & SPO & 890  & 0.81$\pm1.18$    & 0.30$\pm0.43$   & 0.15  & LF \\
20160906\_065905 & 62.4 & -0.8 & HT & SPO & 890  & 56.86$\pm470.80$ & 0.08$\pm0.20$   & 0.02 & LF \\
20161009\_064237 & 32.5 & 2.7  & JF & SPO & 3190 & 3.47$\pm3.16$    & 0.10$\pm0.11$   & 0.06 & LF \\
20161022\_023854 & 54.0 & 1.2  & HT & SPO & 890  & 1.94$\pm3.06$    & 0.22$\pm0.33$   & 0.11 & LF \\
20161023\_022857 & 39.9 & 1    & HT & CTA & 890  & 17.41$\pm27.58$  & 0.05$\pm0.07$   & 0.02 & LF \\
20161026\_023039 & 25.4 & 3.8  & AC & SPO & 4200 & 0.03$\pm0.03$    & 18.71$\pm9.68$  & 18.70 & LF* \\
20161105\_084501 & 65.2 & -0.6 & HT & ORI & 890  & 1.31$\pm1.93$    & 0.13$\pm0.19$   & 0.06 & LF \\
20170228\_092048 & 63.5 & 1.1  & HT & SPO & 890  & 13.73$\pm97.20$  & 0.04$\pm0.06$   & 0.01 & LF \\
20170329\_022149 & 13.0 & 3.1  & AC & SPO & 4200 & 17.89$\pm22.97$  & 0.10$\pm0.14$   & 0.06 & LF \\
20170329\_044039 & 26.7 & 2.9  & JF & SPO & 3190 & 0.22$\pm0.22$    & 0.28$\pm0.34$   & 0.16 & LF \\
20170729\_075956 & 42.2 & 2.2  & JF & SDA & 3190 & 13.73$\pm26.24$  & 0.28$\pm0.31$   & 0.16 & LF \\
20170730\_075821 & 41.0 & 2.4  & JF & SDA & 3190 & 0.28$\pm0.26$    & 0.39$\pm0.44$   & 0.23 & LF \\
20170803\_075755 & 40.2 & 2.4  & JF & SDA & 3190 & 0.07$\pm0.06$    & 2.77$\pm3.01$   & 1.62 & LF* \\ \hline
20150716\_052934 & 41.7 & 2.9  & HT & SPO & 890  & 0.65$\pm1.95$    & 0.85$\pm1.35$   & 0.38 & SB \\ 
20150915\_084106 & 68.8 & 0.6  & HT & SPO & 890  & 0.80$\pm7.62$    & 27.84$\pm22.38$ & 22.66 & SB* \\ \hline
20160709\_042521 & 15.3 & 3.0  & JF & SPO & 3190 & 0.03$\pm0.03$    & 47.27$\pm15.71$ & 43.34 & F* \\
20160710\_065147 & 41.6 & 1.9  & HT & NZC & 890  & 1.15$\pm1.66$    & 6.18$\pm4.42$   & 5.65 & F* \\
20160803\_073937 & 19.3 & 3.5  & AC & SPO & 4200 & 0.16$\pm0.14$    & 31.98$\pm9.59$  & 29.88 & F* \\
20161105\_030709 & 28.8 & 3.2  & AC & NTA & 4200 & 0.02$\pm0.01$    & 39.10$\pm15.61$ & 33.94 & F \\
20170323\_070356 & 57.0 & 3.0  & HT & SPO & 890  & 0.68$\pm0.98$    & 9.35$\pm8.71$   & 6.45 & F* \\
\enddata
\tablecomments{The initial speed is based on the entire meteoroid ablation profile, and not just the second half of the meteor data. The orbit classification and meteoroid density determination is described in \ref{meteoroid_density}. The Origin column labels the meteors as either sporadic (SPO), or by their meteor shower code. The initial mass and luminous efficiency values were found after searching the entire phase space of meteoroid density, shape factor, and drag coefficient, and are mean values unless specified.} Under NOTE, events with an asterisk describe meteors that were found to be sensitive to position picks, and SB refers to meteors showing single body ablation, while LF describes leading fragment events, and F describes those that show long distinct trails.
\end{deluxetable}

\subsection{Meteor event 20161009\_064237}
\begin{figure*}
\gridline{\fig{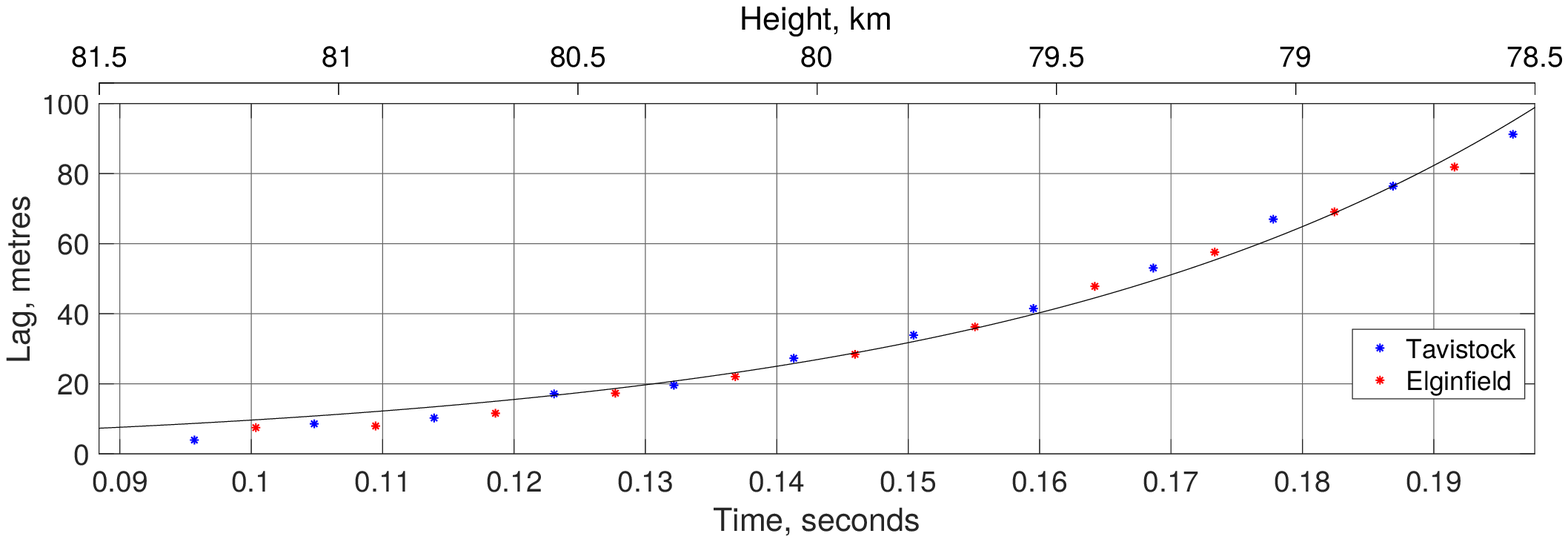}{1\textwidth}{(a)}}
\gridline{\fig{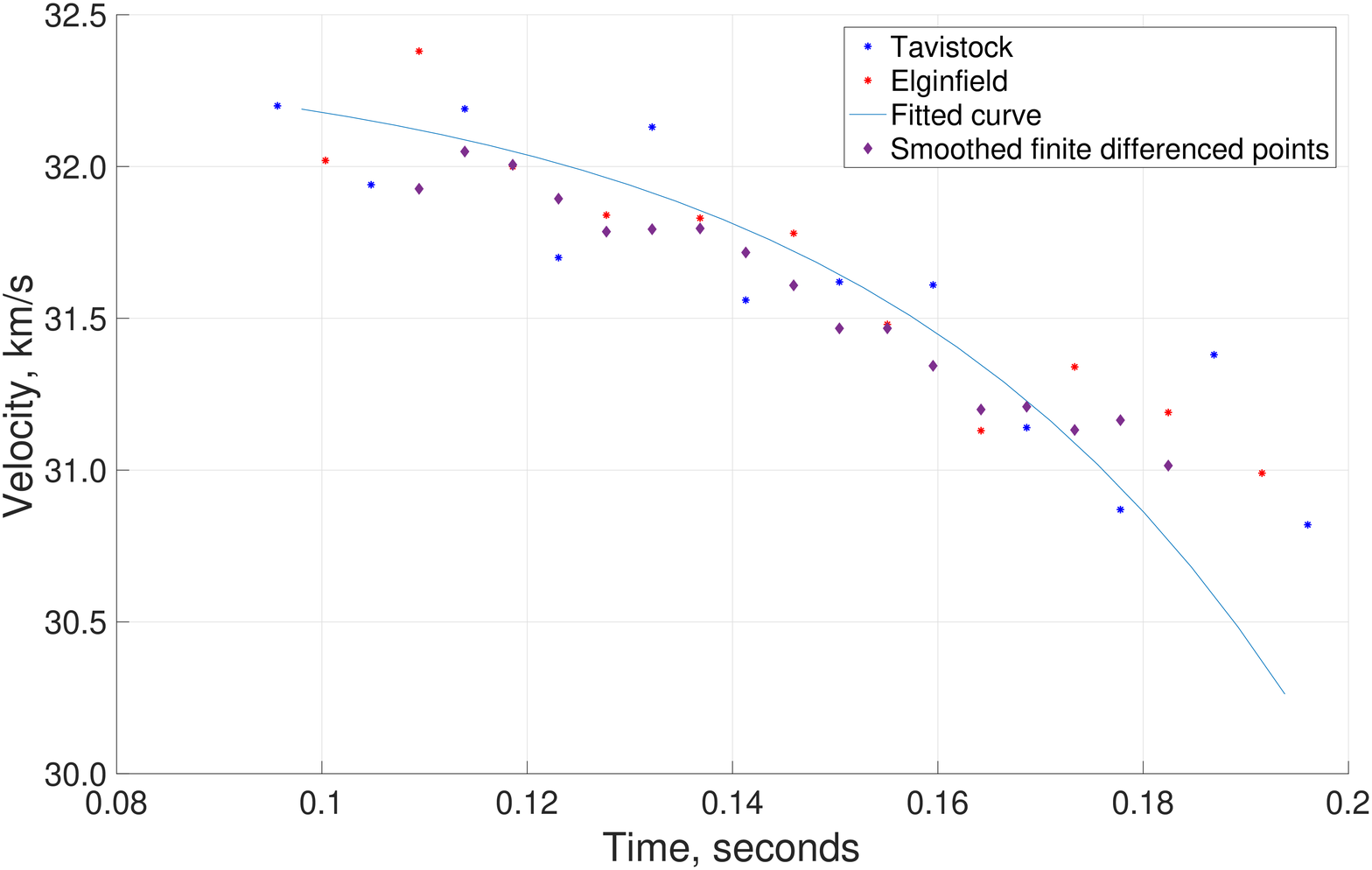}{0.45\textwidth}{(b)}
          \fig{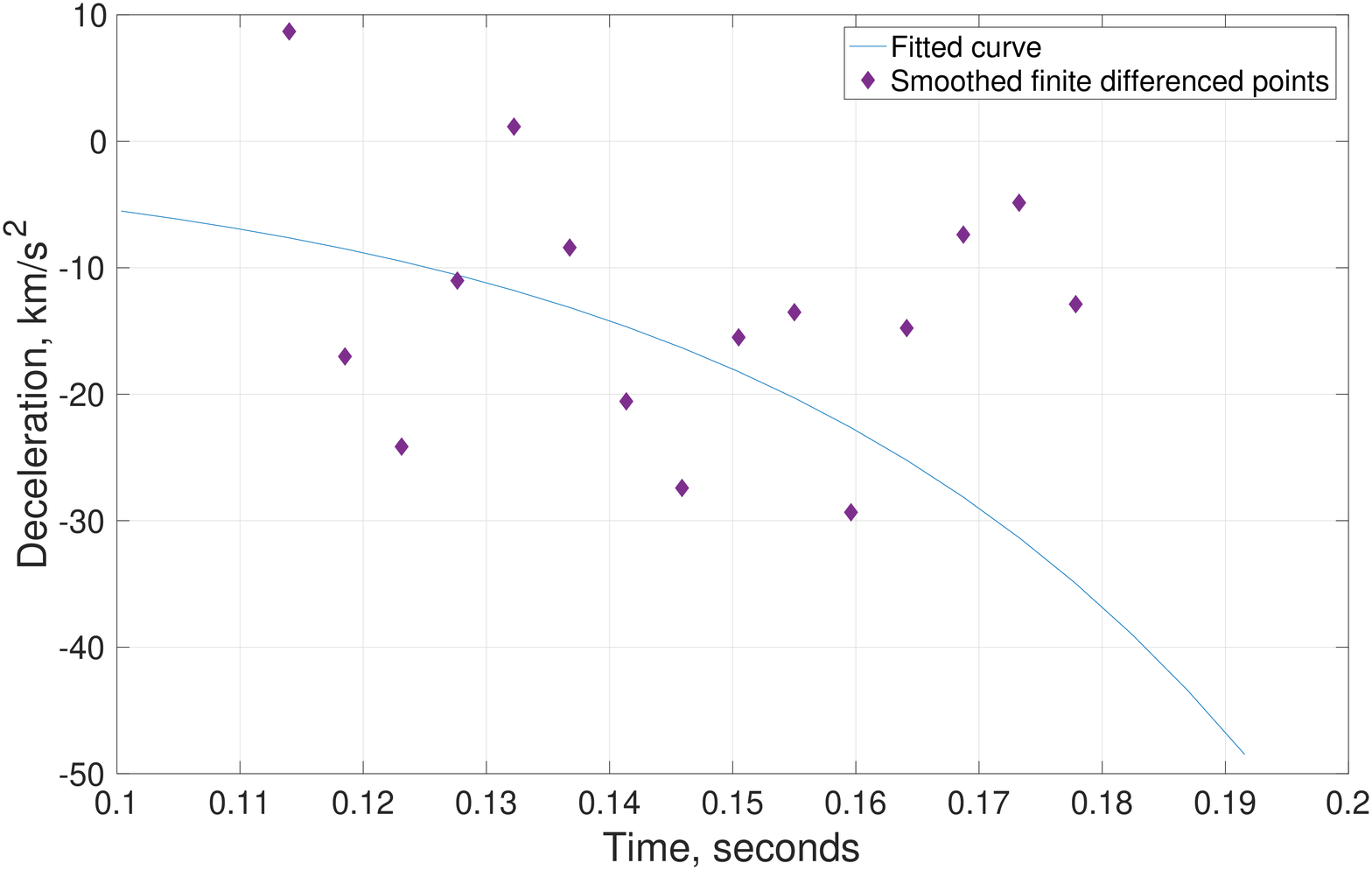}{0.45\textwidth}{(c)}
          }
\gridline{\fig{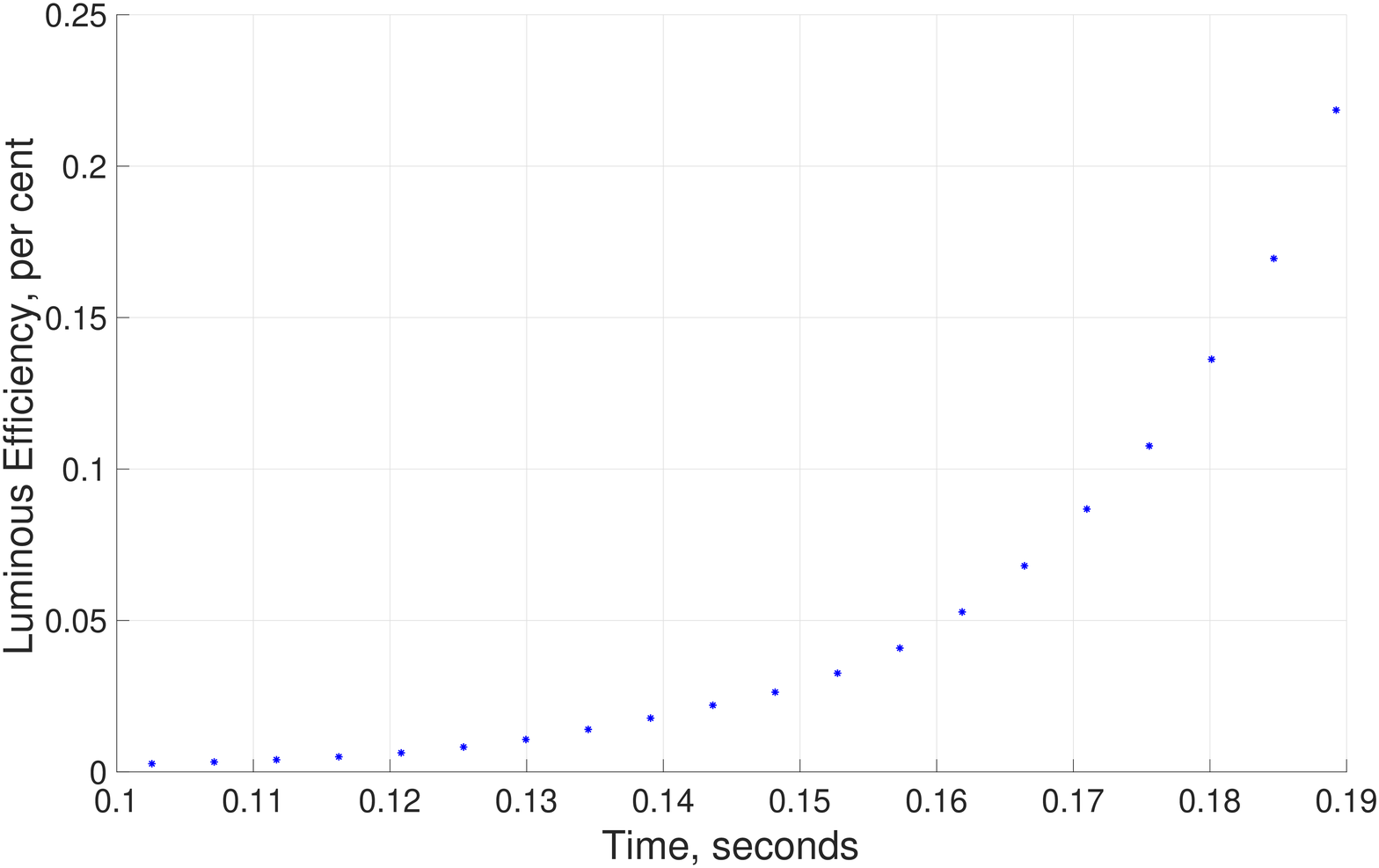}{0.45\textwidth}{(d)}}
\caption{The analysis of meteor event 20161009\_064237. (a) shows the second half of the meteor lag data (with the data points coloured by which station the observation is from), fit with a single-term exponential. (b) shows the speed curve derived from the fitted lag. The red and blue points correspond to the finite differenced values from the two stations. The cyan points are lag data points that are finite differenced over a larger separation to reduce the scatter. (c) shows the deceleration curve and the finite differenced values (also over a larger separation in an attempt to reduce scatter). (d) shows the resulting luminous efficiency profile, based on the speed and deceleration curves, along with an initial meteoroid density of 3190 kgm$^{-3}$; drag coefficient 1; shape factor 1.21.\label{fig:event5}}
\end{figure*}

In Figure \ref{fig:event5} we present the analysis of one event, from fitting the lag, to the final determined luminous efficiency profile. To reduce the noise that comes from finite differencing the values, we smoothed the speed and deceleration values by finite differencing over larger separations. Despite this, there is still considerable scatter in the speed and deceleration points, emphasising the importance of both precise position measurements for this work, and using a suitable fit for the meteor lag. The lag fit parameters for this event are $a=0.8908$ and $b=23.82$, and the following parameters were assumed: meteoroid density of 3190 kgm$^{-3}$; drag coefficient of 1; and shape factor of 1.21. The fitted parameters are based on the original data analysis: they do not take into account searching the parameter space, or the uncertainty of half a pixel in position. The average luminous efficiency over the second half of this meteoroid's ablation was 0.101$\pm$0.111\%.

\subsection{Final Meteor results}
\begin{figure*}
\centering
\includegraphics[angle=90,scale = 0.33]{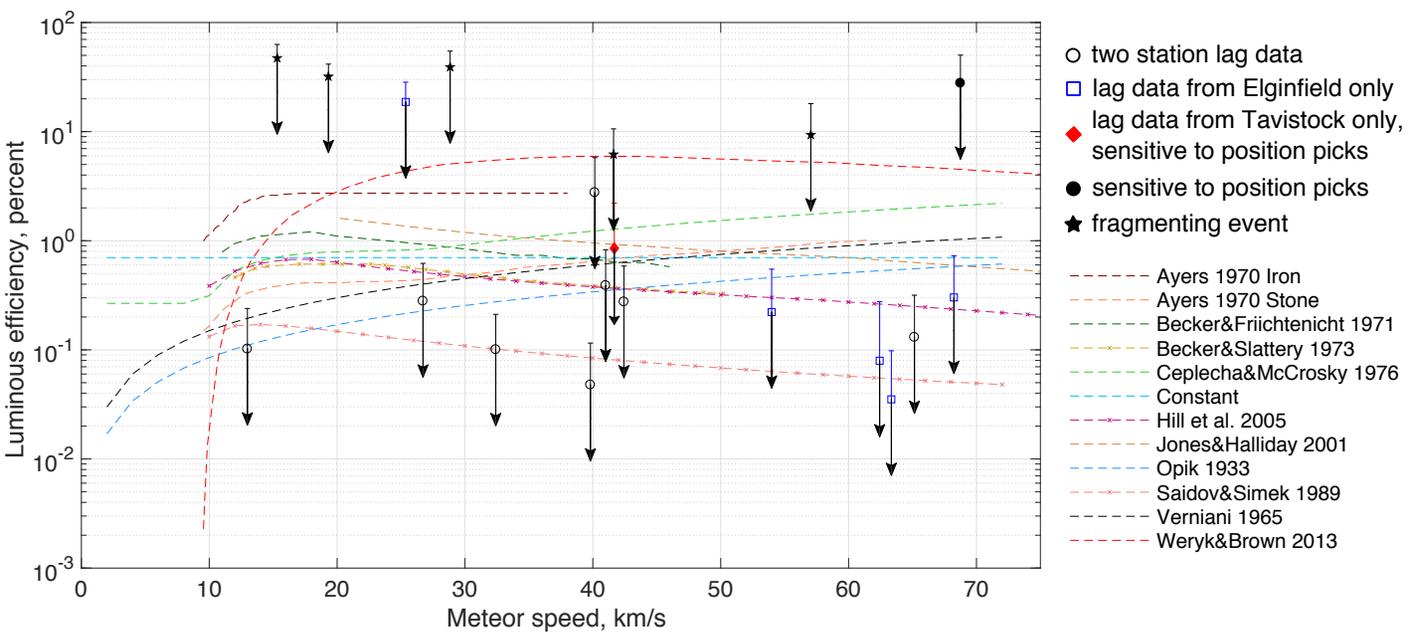}
\caption{Luminous efficiency as a function of initial speed. Note that the initial meteor speed is based on the entire meteoroid ablation profile, and not where the single term exponential begins. The curves showing other studies have not been corrected to a particular bandpass. The instruments used in the \cite{Weryk2013} study are the same ones used in this study, but their results are bolometric values.}
\label{fig:tau_result}
\end{figure*}

The calculated luminous efficiency values for each meteor event analysed, as a function of initial speed, are presented in Figure \ref{fig:tau_result}, plotted over a few past studies for comparison. The error bars are based on searching the entire parameter space of appropriate meteoroid density, drag coefficient, and shape factor, and the downward arrows indicate that the determined luminous efficiency values are upper limits (due to fragmentation), or that the lower bound of the luminous efficiency is less than zero. Some meteor events moved out of the narrow-field camera's field of view at one station: these events are plotted as either blue squares or red diamonds in Figure \ref{fig:tau_result}. The single-term exponential was fit to single station data rather than two station data in those cases; however, data from both stations was used to calculate the meteoroid trajectory.

\subsubsection{Luminous Efficiency and mass}
\begin{figure*}
\plotone{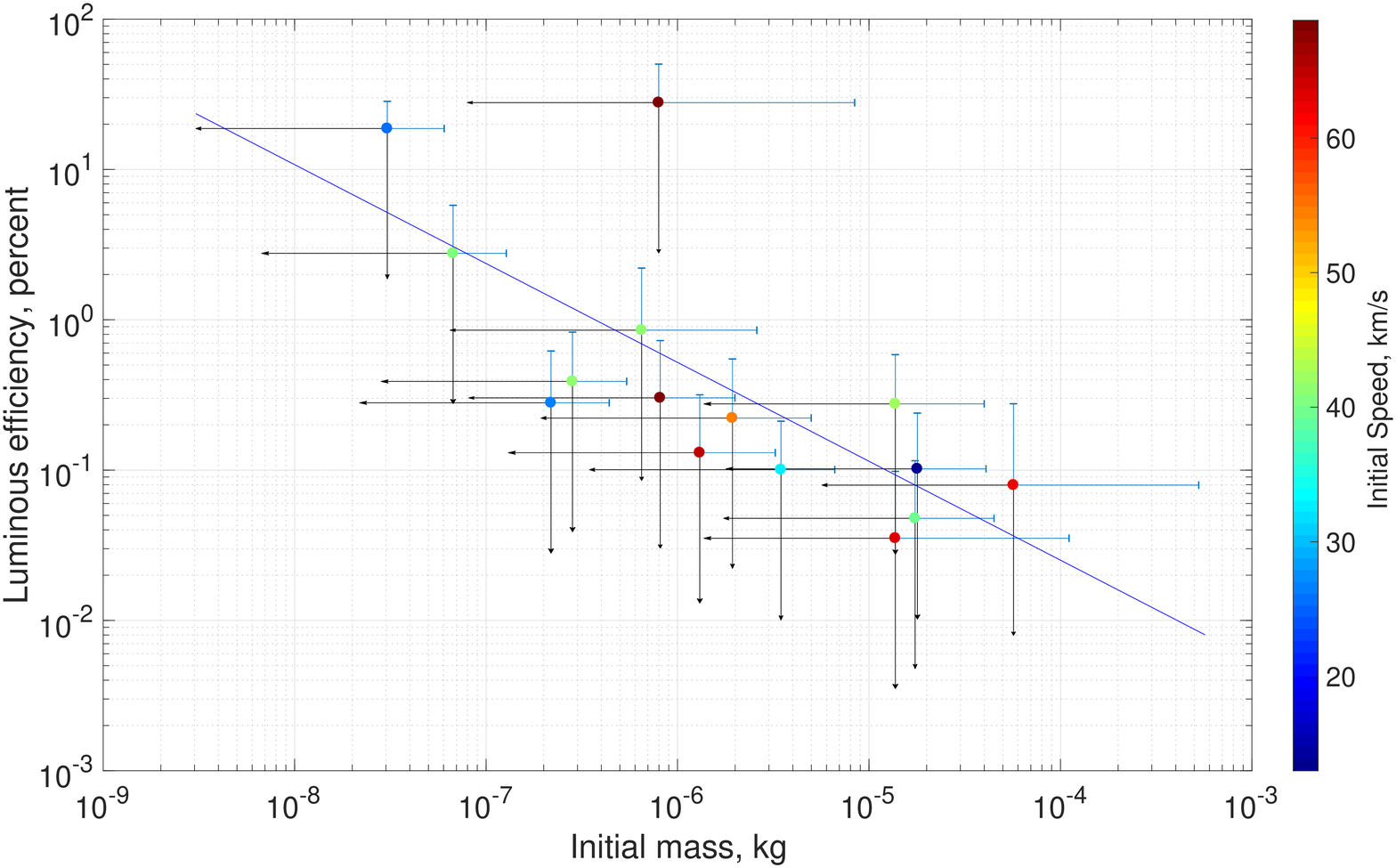}
\plotone{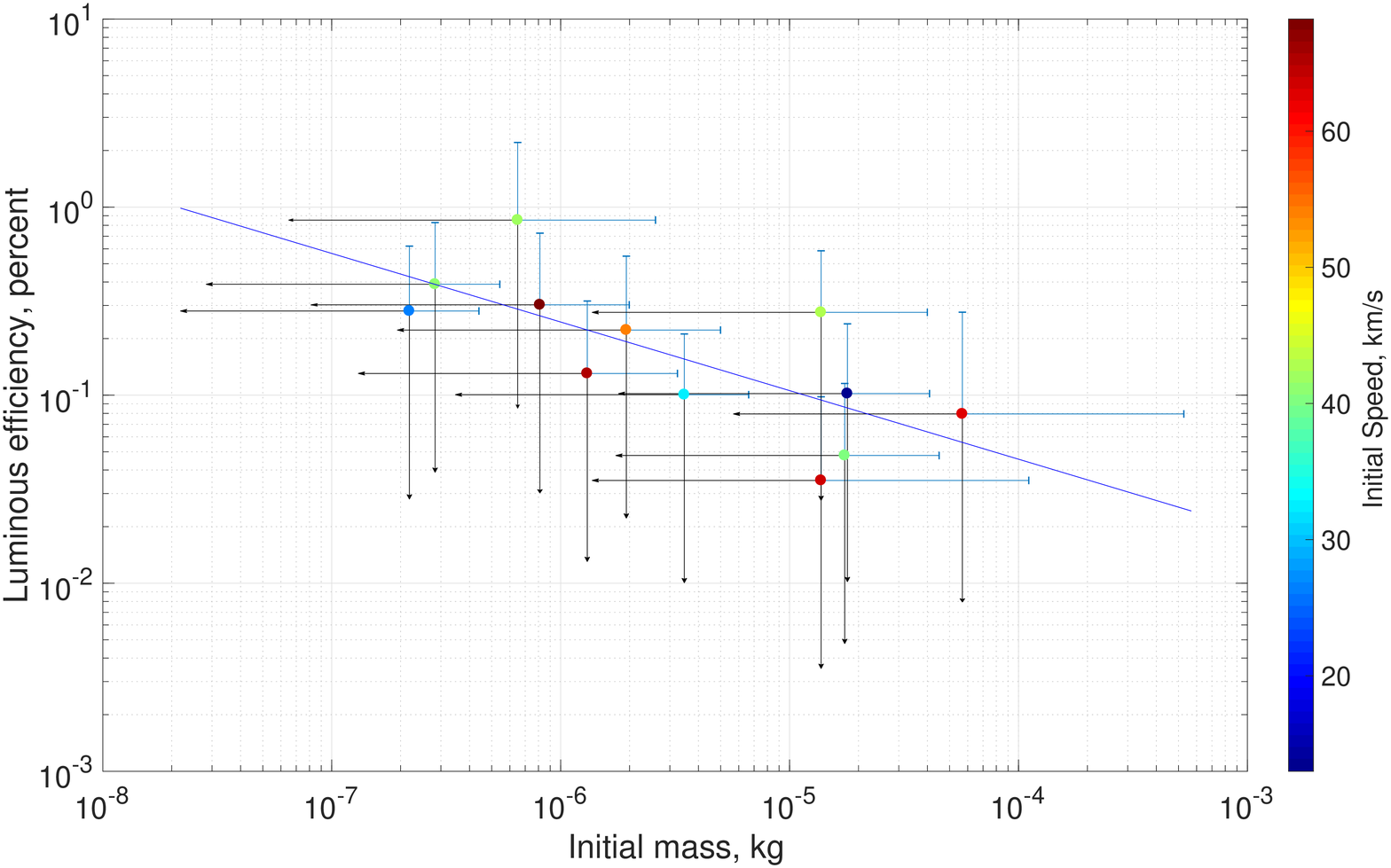}
\caption{Luminous efficiency as a function of initial mass for non-fragmenting events. Each meteor event is coloured according to its initial speed, with the scale given in the colourbar. The upper figure includes all leading fragment and single body events (and the line of best fit has a slope of -0.6578 and an intercept of -4.2303 in log-log space), while the lower figure excluded events from that set that were sensitive to position picks (and has a slope of -0.3647 and an intercept of -2.7994, in log-log space).}
\label{fig:mass}
\end{figure*}

It has been suggested that while the calculated meteoroid mass depends on luminous efficiency, the luminous efficiency may depend on mass according to fireball studies \citep{Halliday1981,Ceplecha1998}. To investigate this, Figure \ref{fig:mass} illustrates the relationship between the average initial mass and luminous efficiency of each meteor. It is worth noting that these initial masses are smaller than the true initial mass of the meteoroid: this is the dynamic mass at the beginning of the second half of the trajectory, and in some cases the mass of the leading fragment instead of the whole meteoroid. As with the above results, these initial masses are an average determined after searching through the entire parameter space. The meteor events are coloured by initial speed. 

\subsubsection{Fragmentation}
Five meteors showing obvious fragmentation (long distinct trails) were analysed with this method. These events served as a sanity check to see whether meteors showing fragmentation will result in unphysical luminous efficiencies, much greater than 100\%. To be able to plot these results, we ignored luminous efficiency values greater than 100\% when computing the average. A clear distinction between the fragmenting and non-fragmenting meteor events can be seen in Figure \ref{fig:tau_result}.

\section{Discussion}
As discussed in \cite{Subasinghe2017}, our method involves fitting the second half of the lag because the luminous efficiency can only be calculated in the part of the trajectory with maximum deceleration. For ideal data, the change in functional fit from a two-term exponential to a single exponent worsened the agreement of values given in Table 1 of \cite{Subasinghe2017} with the 0.7\% constant luminous efficiency used to simulate the events. In spite of this, the single term exponential is a better choice for real data with noise. \newline

To test the effect of the fit on the derived luminous efficiency values, and as an extension of the work done in \cite{Subasinghe2017}, a simple polynomial (lag = ax$^3$ + bx$^2$ + cx + d, which gives a linear deceleration) was fit to meteor event 20161009\_064237. The polynomial could not be of order two or less, as that resulted in a constant deceleration, meaning the dynamic mass was not changing. The luminous efficiency was also determined using a simple point to point method, in which the mean deceleration was determined for two different sections of the meteor data near the end of ablation (resulting in two deceleration values). These were used to find a dynamic mass, compared to the photometric mass lost between those points and used to calculate the luminous efficiency. These two methods found luminous efficiency values that were within a factor of three of the value determined by fitting a single-term exponential function to the second half of the meteor lag data. The exponential fit was used for this analysis as it best describes the atmospheric density that is encountered by the meteoroid.  \newline

We analysed fifteen non-fragmenting meteors, which is a very small fraction of the thousands recorded by CAMO. However, as seen in Figure \ref{fig:frames}, the images for the thirteen leading fragment events show very little to no wake, suggesting fragmentation is not important and that they can be treated as single bodies validating our method and results. Two of the meteor events (20150716\_052934 and 20150915\_084106) analysed show single body ablation (but are not leading fragment events); however, they are likely undergoing fragmentation on a scale that we cannot resolve, as with the CAMO meteor in \citet{CampbellBrown2017} which appeared to have negligible wake but could only be modelled with significant fragmentation. Thus those two results should be considered upper limits on the luminous efficiency for those events - the determined dynamic mass (which considers only the largest fragment) will be less than the photometric mass (which considers light production from all fragments), and for the same amount of light production, this would cause the luminous efficiency to be artificially increased. \newline

Our simulated noise analysis shows that for low (10$^{-6}$ kg), mid (10$^{-5}$ kg), and high-mass meteors (10$^{-4}$ kg), low speed meteors are most likely to produce luminous efficiency results closest to the value used in the simulation. This is likely because slower meteors decelerate more, so uncertainties affect the dynamic mass less than for fast meteors which ablate before they significantly decelerate. For the entire range of meteor speeds, our method tends to underestimate the luminous efficiency of meteors. \newline

\begin{figure}
\plotone{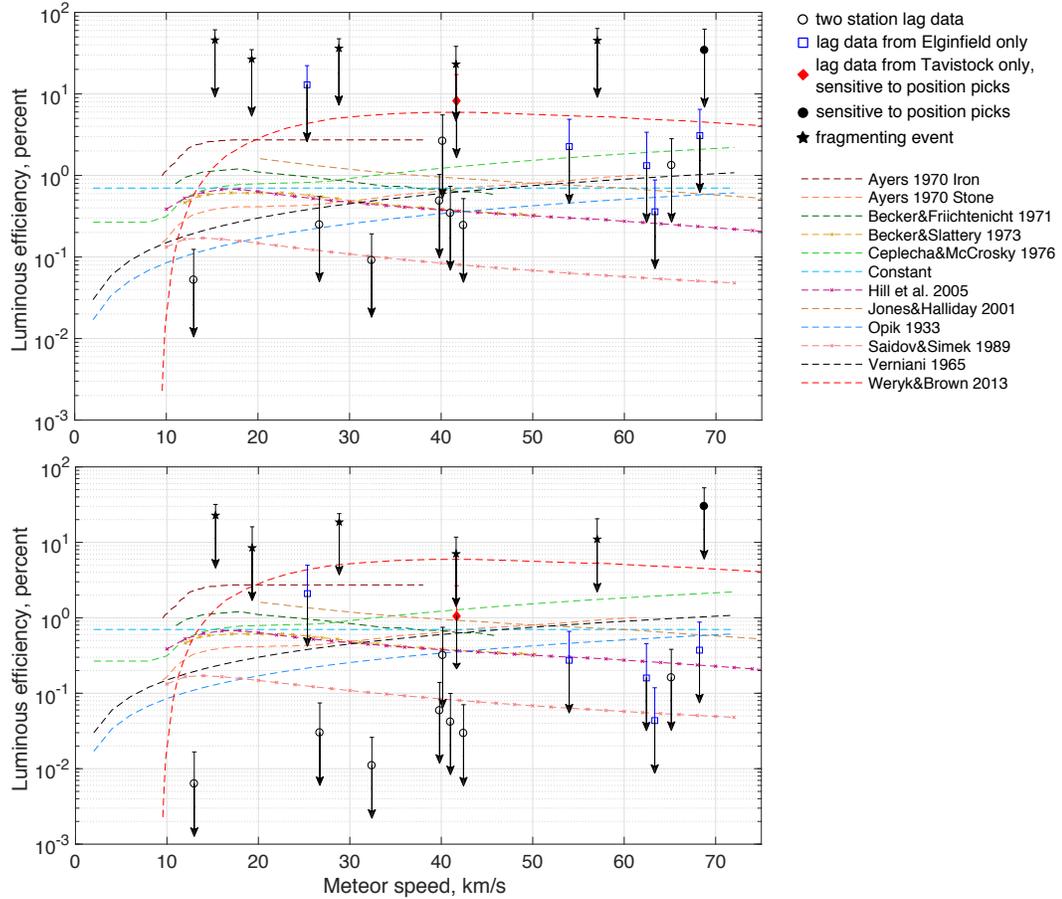}
\caption{Luminous efficiency as a function of initial speed, assuming each meteoroid has a bulk density of 3000 kgm$^{-3}$ (upper plot), or 1000 kgm$^{-3}$ (lower plot).}
\label{fig:tau_1000}
\end{figure}

We used meteoroid bulk density values from \cite{Kikwaya2011a} based on the Tisserand parameters of our meteor events. The results from \cite{Kikwaya2011a} assumed a luminous efficiency value. To eliminate any bias this assumed luminous efficiency value may have had on our derived luminous efficiency results, we present the derived luminous efficiency results with meteoroid density 1000 kgm$^{-3}$ and 3000 kgm$^{-3}$ for each event, in Figure \ref{fig:tau_1000}. Assuming a density of 1000 kgm$^{-3}$ has the effect of decreasing the calculated luminous efficiencies for slow meteors, while not significantly changing those of faster meteors, while assuming a density of 3000 kgm$^{-3}$ increases the calculated luminous efficiencies of fast meteors, while not significantly changing those of slower meteors. Both assumed densities show an increase in calculated luminous efficiency with speed, while the density values from \cite{Kikwaya2011a} and \cite{Kikwaya2011} show a constant relationship with speed.\newline

The main results of this work are presented in Figures \ref{fig:tau_result} and \ref{fig:mass}. Our results are typically consistent with lower values of luminous efficiency from previous studies, and seem to rule out the highest luminous efficiencies. It is difficult to directly compare our results to others because several past studies knew the composition of their meteoroids (in the case of lab meteoroids, or artificial meteoroids). We attempt to account for this by exploring a large range of possible meteoroid densities. \newline

Figure \ref{fig:mass} shows that there is a relationship between luminous efficiency and initial mass, which is not related to meteoroid speed. Unexpectedly, it shows that meteoroids with smaller mass radiate light more efficiently than more massive meteoroids. The first plot in Figure \ref{fig:mass} shows the linear fit (in log-log space) when including all the analysed meteor events, and the second figure excludes the three events that were found to be very sensitive to meteor position picks. The entire parameter space of drag coefficient, meteoroid density, shape factor, and random errors in position was searched, with random errors of up to half a pixel being added. Three of the fifteen events were found to be very sensitive to these position pick variations, leading to wildly different luminous efficiencies for similar position picks. Even when these three events are removed, a negative linear trend is still apparent in the luminous efficiency-mass plot. The uncertainty in each point is large, however, and more meteor events may cause this to change. \newline

It is important to keep in mind the low number of meteor events studied here. While Figure \ref{fig:mass} shows negative linear trends, these events are not necessarily representative of the entire population of meteors observed with the CAMO system. There are no meteors in the low luminous efficiency - low mass area of the plots in Figure \ref{fig:mass}: it is unlikely that the CAMO system could observe such faint meteors. However, there are also no high luminous efficiency - high mass meteors: the CAMO system should be able to see these bright meteors. 
One matter of great interest is the dependence of luminous efficiency on speed, since some previous studies predict an increase \citep[e.g.][]{Verniani1965}, and some a decrease with speed \citep[e.g][]{Hill2005}. Our results using the densities of \citet{Kikwaya2011a} show no clear trend with speed, while using a constant density there is a weak increase with speed. At low speeds, our results are even less conclusive; we have a single meteor event with an initial speed less than 20 kms$^{-1}$, which does not allow us to corroborate or reject the steep increase in luminous efficiency at low speeds typically found in past studies. The error in each event is typically greater than the derived luminous efficiency value, which makes it unreasonable to draw lower bounds on the luminous efficiency value with this study. This is indicated by downward pointing arrows. It is not unreasonable to assume that the trend in luminous efficiency with speed depends on the composition, with some atoms radiating more effectively at higher collision energies, and some less effectively. In this case, any trend will be masked if the meteoroids have different compositions.\newline

An important note is that thirteen of our fifteen meteor events were leading (or terminal) fragments. These fragments were composed of the strongest material in the meteoroid, which is implied by the fact that they continued to ablate after most of the meteoroid had ablated. These fragments may have had a different composition (and therefore spectrum) from the rest of the meteoroid, which means they may have a different luminous efficiency than their parent meteoroid. It is therefore difficult to compare these leading fragment meteor events to other studies. For example, if the leading fragments contain little volatile sodium, they would produce less light than sodium-rich parts of the meteoroid.

\section{Conclusion}
This paper presents the most recent study of modern high-resolution meteor observations used to determine luminous efficiency, by comparing the dynamic and photometric masses. The second half of the observed meteor lag is fit with a single-term exponential, and the resulting speed and deceleration curves are used in conjunction with best fit values for meteoroid density, shape factor, and drag coefficient. To determine uncertainties, the entire phase space of potential values for the above parameters was searched. Fifteen meteor events were observed, thirteen of which displayed leading fragment behaviour (a fragment that persisted after the majority of the meteoroid had ablated, and showed essentially no fragmentation), and two which show as close to single body ablation as we could find. While there is no obvious relationship found between luminous efficiency and initial meteor speed, this may be due to the variability of meteoroid compositions, on which we have gathered no information for this study. It is also difficult to directly compare our results to past studies (artificial meteoroids, lab studies, simultaneous radar/optical studies) as our leading fragment events are composed of the strongest material in each meteoroid, which may not necessarily be represented in past studies. There seems to be an unexpected negative linear relationship between luminous efficiency and initial meteoroid mass; however, as there are only fifteen events, this may not be meaningful. In the future, we will add spectral capabilities to the CAMO system and collect more data with the necessary high-resolution calibrations. This may cause relationships between luminous efficiency and other parameters (such as mass and speed) to reveal themselves.

\acknowledgments
This work was supported by the NASA Meteoroid Environment Office [grant NNX11AB76A]. DS thanks the province of Ontario for funding. Thanks to Z. Krzeminski, J. Gill and D. Vida for their assistance with data collection and reduction. The authors thank an anonymous reviewer for their helpful comments.
\software{ASGARD \citep{Weryk2008}, METAL \citep{Weryk2012}, mirfit}

\bibliography{tau_two}

\end{document}